\documentclass[twocolumn,tighten]{aastex63}

\usepackage{amsmath}
\usepackage{color}
\usepackage{multirow}
\usepackage{hyperref}
\usepackage{bbm}
\AtBeginDocument{\renewcommand{\d}{\textrm{d}}}
\DeclareMathOperator\erf{erf}

\shorttitle{Can the Cosmological Dilation Explain the Skewness in the Gamma-Ray Burst Duration Distribution?}
\begin{document}

\title{Can the Cosmological Dilation Explain the Skewness in the Gamma-Ray Burst Duration Distribution?}

\author[0000-0003-4666-0154]{Mariusz Tarnopolski}
\email{mariusz.tarnopolski@uj.edu.pl}
\affiliation{Astronomical Observatory, Jagiellonian University, Orla 171, 30--244, Krak\'ow, Poland}

\begin{abstract}
In order to explain the origin of skewness in the gamma-ray burst (GRB) duration distribution, a statistical model, based on the cosmological time dilation and taking into account the GRB formation rate and instrumental effects, is examined. It is concluded that the considered effects are able to account for only a small fraction of the observed skewness. Therefore, its origin needs to be searched for in the processes governing the progenitors of GRBs.​
\end{abstract}

\keywords{Gamma-ray bursts --- Cosmology --- Astronomy data analysis --- Astrostatistics}

\section{Introduction}
\label{introduction}

Gamma-ray bursts (GRBs; \citealt{klebesadel}) are commonly divided into two classes: short (attributed to compact-object mergers) and long (massive-star collapsars). The division is clearly visible in the bimodal distribution of durations $T_{90}$ (i.e., time during which 90\% of the GRB's fluence is detected; its range spans $\sim 6$ orders of magnitude, from milliseconds to thousands of seconds), and occurs at $T_{90}\simeq 2\,{\rm s}$ (\citealt{kouve93}; but see also \citealt{bromberg13,tarnopolski15a}). Since a putative third, intermediate-duration class was reported \citep{horvath98}, the distribution was routinely modeled with a mixture of normal distributions in several subsequent works (e.g., \citealt{horvath02,horvath08,zhang08,huja09,horvath10,zhang16}), which often concluded that a third component is required to fit the data appropriately, and attributed physical meaning to it (although see also \citealt{bhat,kienlin20} for an analysis of the Fermi/GBM team that does not find evidence for a third class).

However, the third component need not be evidence of a physically motivated group, but a sign of inherent skewness of the long-GRB class \citep{koen,tarnopolski15b}. Indeed, when modeling with skewed distributions, only two components are required to model the data appropriately \citep{tarnopolski16a,tarnopolski16c,kwong}, implying the third Gaussian one is spurious, and appears because of modeling an intrinsically skewed distribution with symmetric ones.\footnote{On the other hand, it was suggested that this elusive group might be attributed to X-ray flashes \citep{veres10,grupe13}, but no decisive conclusions could be formulated \citep{ripa14,ripa16}.}

Similar conclusions were drawn from investigating the two-dimensional space spanned by hardness ratios and durations: Gaussian mixtures often pointed at three groups \citep{horvath06,ripa09,horvath10,veres10,horvath18}, but some works indicated that only two are required \citep{ripa12,yang16,tarnopolski19b,kienlin20}. Moreover, considering skewed distributions, two-component mixtures were also pointed at \citep{tarnopolski19a,tarnopolski19b}. In higher-dimensional spaces things become less unambiguous, though \citep{mukherjee,chattopadhyay07,chattopadhyay17,chattopadhyay18,acuner,modak18,horvath19,toth19,tarnopolski19c}. For the most recent overview on the topic of GRB classes, see \citep{tarnopolski19c}.

Finally, the rest-frame durations (possible to derive for GRBs with measured redshift $z$) appear to be sufficiently well described with a two-component Gaussian mixture \citep{huja09,tarnopolski16c,zhang16,zitouni18}. This fact, coupled with skewness in the observer frame and cosmological distances (up to $z\gtrsim 9$), suggests that convolution with the redshift distribution might transform the intrinsically Gaussian durations into a skewed distribution \citep{tarnopolski19a}. The aim of the presented work is to investigate this hypothesis.

This paper is structured as follows. Section~\ref{methods} describes the utilized samples, their properties, the employed probability distributions, and the star formation rates (SFRs) that the GRBs follow. Section~\ref{model} outlines the investigated statistical model. In Sect.~\ref{results} results of the analyses are displayed. Section~\ref{discussion} is devoted to discussion and interpretation. Concluding remarks are gathered in Sect.~\ref{summary}. 

\section{Data Sets and Methods}
\label{methods}

\subsection{Samples}
\label{samples}

The Fermi/GBM catalogue\footnote{\url{https://heasarc.gsfc.nasa.gov/W3Browse/fermi/fermigbrst.html}, accessed on 2020 February 10.} \citep{gruber,kienlin,bhat,kienlin20} contains 2733 GRBs with measured $T_{90}$ (from GRB 080714086 to GRB 200208052), among which are 449 short and 2284 long ones. \citet{zitouni18} compiled the redshifts of 134 events (11 short and 123 long). 

Swift/BAT\footnote{\url{https://swift.gsfc.nasa.gov/archive/grb\_table}, accessed on 2020 February 10} \citep{gehrels} observed 1264 GRBs (from GRB 041217 to GRB 200205B), among which 118 are short, and 1146 are long. The subsample of GRBs with a measured redshift consists of 366 events (11 short and 355 long). Redshifts in both Fermi and Swift samples extend to $z_{\rm max}=8$.

The compilation of \citet{wang20}, who gathered the redshifts\footnote{Three GRBs have $z>8$.} and durations of 568 GRBs, is employed hereinafter to investigate the distribution of intrinsic durations.

\subsection{Distributions}
\label{distributions}

The probability density function (PDF) of a standard normal distribution, $\mathcal{N}(0,1)$, is denoted by $\varphi(x) = 1/\sqrt{2\pi} \exp(-x^2/2)$, and its cumulative distribution function is $\Phi(x) = \int_{-\infty}^x \varphi(t) \d t = \frac{1}{2} \left[ 1+\erf( x/\sqrt{2}) \right]$. The PDF of a general normal distribution, $\mathcal{N}(\mu,\sigma^2)$, is therefore
\begin{equation}
f^{(\mathcal{N})}(x) = \frac{1}{\sigma} \varphi\left(\frac{x-\mu}{\sigma}\right).
\label{eq1}
\end{equation}

The PDF of a truncated normal distribution, $\mathcal{TN}_{(0,+\infty)}(\mu,\sigma^2)$, with support in $(0,+\infty)$, is
\begin{equation}
f^{(\mathcal{TN})}(x) = 
\begin{cases}
\begin{split}
\frac{f^{(\mathcal{N})}(x)}{\Phi\left(\frac{\mu}{\sigma}\right)}, & \quad x\in(0,+\infty) \\
0, & \quad x\notin(0,+\infty)
\end{split}
.
\end{cases}
\label{eq2}
\end{equation}

A skew normal distribution, $\mathcal{SN}(\mu,\sigma^2,\lambda)$ \citep{azzalini85}, is defined via its PDF to be
\begin{equation}
f^{(\mathcal{SN})}(x) = \frac{2}{\sigma}\varphi\left(\frac{x-\mu}{\sigma}\right)\Phi\left(\lambda\frac{x-\mu}{\sigma}\right).
\label{eq3}
\end{equation}
When $\lambda=0$, the $\mathcal{SN}(\mu,\sigma^2,\lambda)$ distribution reduces to $\mathcal{N}(\mu,\sigma^2)$. Skewness of the $\mathcal{SN}(\mu,\sigma^2,\lambda)$ distribution is $\gamma = \frac{4-\pi}{2} \left(\delta\sqrt{2/\pi}\right)^3 \left(1-2\delta^2/\pi\right)^{-3/2}$, where $\delta=\lambda/\sqrt{1+\lambda^2}$.

An extension of the $\mathcal{SN}(\mu,\sigma^2,\lambda)$ distribution, denoted $\mathcal{SNE}(\mu,\sigma^2,\lambda,\xi)$ \citep{azzalini85,henze}, is given as
\begin{equation}
f^{(\mathcal{SNE})}(x) = \frac{\varphi\left( \frac{x-\mu}{\sigma} \right) \Phi\left( \lambda\frac{x-\mu}{\sigma}+\xi \right)}{\sigma\Phi\left( \frac{\xi}{\sqrt{1+\lambda^2}} \right)}.
\label{eq4}
\end{equation}
When $\xi=0$, the $\mathcal{SNE}$ distribution becomes an $\mathcal{SN}$ distribution from Eq.~(\ref{eq3}). When $\lambda=0$, the $\mathcal{SNE}$ distribution becomes independent on $\xi$, and reduces to a normal distribution.

The $\mathcal{SN}$ distribution can be represented \citep{henze} in terms of normal and truncated normal distributions:
\begin{equation}
\lambda|U|+V\sim \mathcal{SN}(0,\sqrt{1+\lambda^2},\lambda),
\label{eq5}
\end{equation}
where $U,V$ are independent standard normal random variables; the distribution of $|U|$ is half-normal, i.e. $|U| \sim \mathcal{TN}_{(0,+\infty)}(0,1)$. Equation~(\ref{eq5}) generalizes to an $\mathcal{SNE}$ distribution (Equation~(\ref{eq4})) when a sum of general $\mathcal{N}(a,b^2)$ (Equation~(\ref{eq1})) and $\mathcal{TN}_{(0,+\infty)}(c,d^2)$ (Equation~(\ref{eq2})) random variables is considered (see Appendix~\ref{appA} for a sketch of the derivation). The $\mathcal{SNE}$ distribution arises naturally when the half-normal distribution in the stochastic representation in Eq.~(\ref{eq5}) is allowed to be in a more general form of the $\mathcal{TN}$ distribution from Eq.~(\ref{eq2}). Such form is apparently more suitable to model the redshift distribution (see, however, the discussion in Sect.~\ref{disc::5.1}). 

\subsection{Fitting the Duration Distributions}
\label{fitting}

The logarithms of observed durations, $\log T_{90}^{\rm obs}$, were fitted with mixtures of two and three components of $\mathcal{N}$ or $\mathcal{SN}$ distributions by maximizing the log-likelihood. Based on the Akaike and Bayesian information criteria, both Fermi and Swift GRBs are well described by a mixture of two $\mathcal{SN}$ distributions. Skewness of the long-GRB component is $\gamma = -0.20$ for Fermi ($\lambda=-1.2$), and $\gamma = -0.63$ for Swift ($\lambda=-2.8$). See \citet{tarnopolski16a} for additional details on the fitting.

Similarly the intrinsic durations from the sample of \citet{wang20} were examined. The mean and standard deviation of the long-GRB group, in the two-component Gaussian mixture that was used to model $\log T_{90}^{\rm int}$, are 1.26 and 0.49, respectively. Note that a mixture with two $\mathcal{SN}$ components is a competing fit, although from the argument of simplicity the Gaussian mixture is preferred.

\subsection{SFRs}
\label{sfrs}

\begin{figure}
\includegraphics[width=\columnwidth]{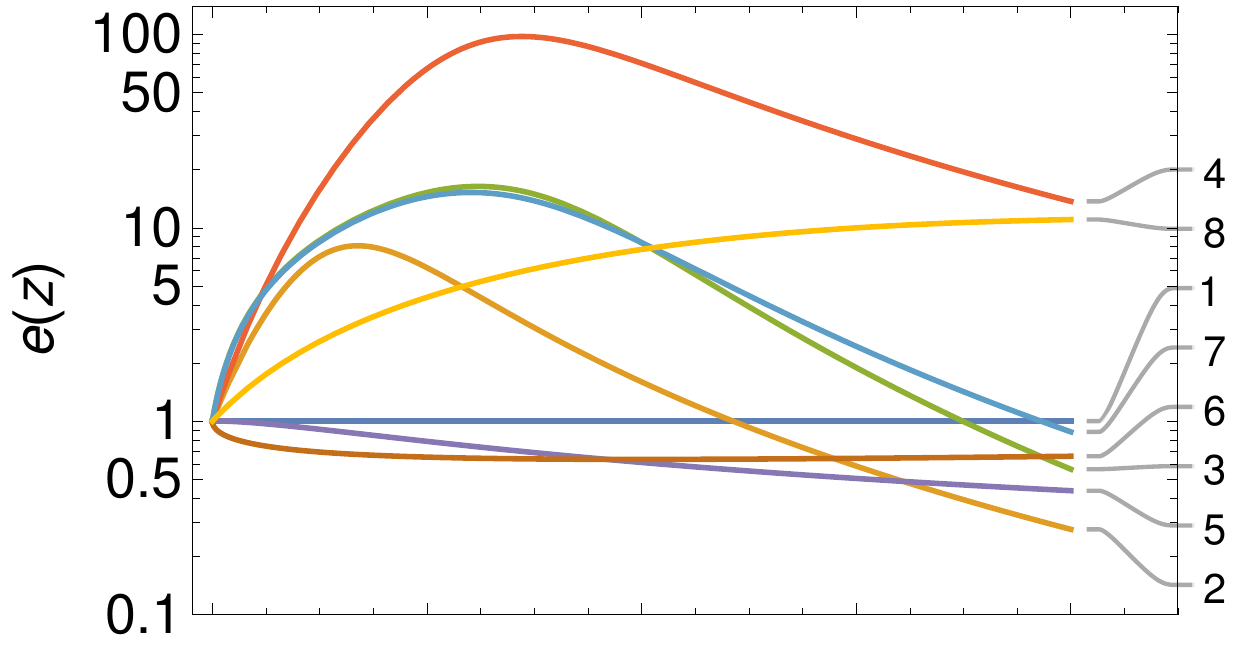}
\includegraphics[width=\columnwidth]{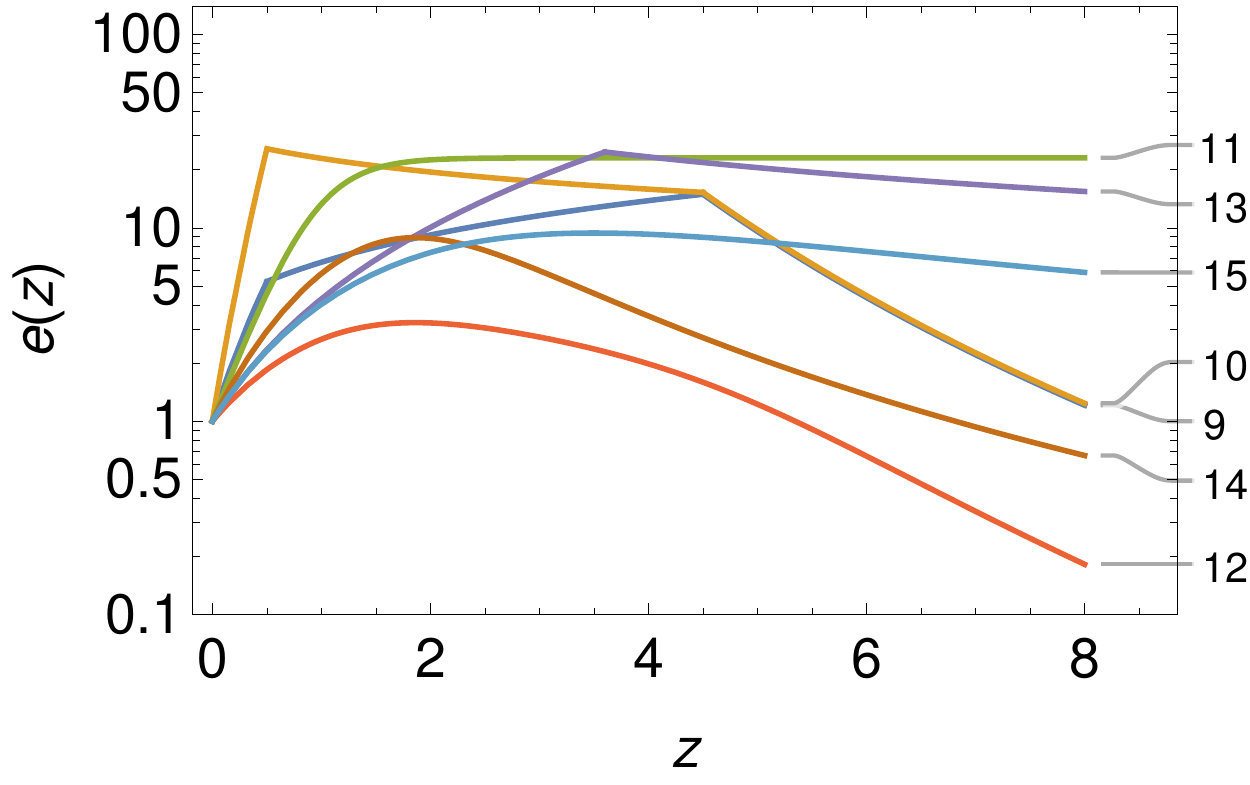}
\caption{The scaled SFRs. Labels correspond to Table~\ref{table::sfrs}. }
\label{fig0}
\end{figure}

Several parametric formulations for the SFR have been considered in the literature, and fitting different data sets lead to different results. Table~\ref{table::sfrs} gathers the 15 SFRs, represented via the scaled $e(z)$ (see Sect.~\ref{model}), that are utilized hereinafter. They are also displayed in graphical form in Fig.~\ref{fig0}.

\begin{deluxetable*}{cclccc}
\tabletypesize{\footnotesize}
\tablecolumns{6}
\tablewidth{0pt}
\tablecaption{Formulas for the Scaled SFRs, and Skewnesses of the Resulting Redshift Distributions. \label{table::sfrs}}% also check \citep{lapi17} 
\tablehead{
\colhead{Label} & \colhead{$e(z)$} & \colhead{Parameters} & \colhead{Skew. of $P_0(z)$} & \colhead{Skew. of $P_{\rm Swift}(z)$} & \colhead{Skew. of $P_{\rm Fermi}(z)$} }
\startdata
1 & 1 & & $-0.084$ & $0.464$ & $0.182$ \\
\hline
2 & $\frac{1+a_1}{(1+z)^{-a_2}+a_1(1+z)^{a_3}}$ & $\begin{array} {lcl} a_1 &=& 0.005 \\ a_2 &=& 3.3 \\ a_3 &=& 3.0 \end{array}$ & $0.499$ & $0.290$ & $0.415$ \\
\hline
3 & $\frac{1+\frac{a_2z}{a_1}}{1+\left( \frac{z}{a_3} \right)^{a_4}}$ & $\begin{array} {lcl} a_1 &=& 0.015 \\ a_2 &=& 0.1 \\ a_3 &=& 3.4 \\ a_4 &=& 5.5 \end{array}$ & $-0.066$ & $0.008$ & $-0.092$ \\
\hline
4 & $\frac{1+a_1}{(1+z)^{-a_2}+a_1(1+z)^{a_3}}$ & $\begin{array} {lcl} a_1 &=& 0.0001 \\ a_2 &=& 4.0 \\ a_3 &=& 3.0 \end{array}$ & $-0.013$ & $-0.128$ & $-0.062$ \\
\hline
5 & $\frac{1+\frac{a_2z}{a_1}}{1+\left( \frac{z}{a_3} \right)^{a_4}}$ & $\begin{array} {lcl} a_1 &=& 0.015 \\ a_2 &=& 0.12 \\ a_3 &=& 3.0 \\ a_4 &=& 1.3 \end{array}$ & $0.112$ & $0.464$ & $0.304$ \\
\hline
6 & $\frac{1+\frac{a_2z}{a_1}}{1+\left( \frac{z}{a_3} \right)^{a_4}}$ & $\begin{array} {lcl} a_1 &=& 0.011 \\ a_2 &=& 0.12 \\ a_3 &=& 3.0 \\ a_4 &=& 0.5 \end{array}$ & $-0.059$ & $0.459$ & $0.250$ \\
\hline
7 & $\frac{1+\frac{a_2z}{a_1}}{1+\left( \frac{z}{a_3} \right)^{a_4}}$ & $\begin{array} {lcl} a_1 &=& 0.0157 \\ a_2 &=& 0.118 \\ a_3 &=& 3.23 \\ a_4 &=& 4.66 \end{array}$ & $-0.001$ & $0.051$ & $-0.028$ \\
\hline
8 & $\frac{1+a_1}{(1+z)^{-a_2}+a_1(1+z)^{a_3}}$ & $\begin{array} {lcl} a_1 &=& 0.005 \\ a_2 &=& 4.5 \\ a_3 &=& 1.0 \end{array}$ & $-0.515$ & $0.219$ & $0.039$ \\
\hline
9$^a$ & $ \begin{cases} a_0(1+z)^\alpha & 0\leqslant z\leqslant z_1 \\ b_0(1+z)^\beta & z_1 < z\leqslant z_2 \\ c_0(1+z)^\gamma & z_2 < z \end{cases} $ & $\begin{array} {lcl} \alpha &=& 4.1 \\ \beta &=& 0.8 \\ \gamma &=& -5.1 \\ z_1 &=& 0.5 \\ z_2 &=& 4.5 \end{array}$ & $-0.336$ & $0.209$ & $-0.155$ \\
\hline
10$^a$ & $ \begin{cases} a_0(1+z)^\alpha & 0\leqslant z\leqslant z_1 \\ b_0(1+z)^\beta & z_1 < z\leqslant z_2 \\ c_0(1+z)^\gamma & z_2 < z \end{cases} $ & $\begin{array} {lcl} \alpha &=& 8.0 \\ \beta &=& -0.4 \\ \gamma &=& -5.1 \\ z_1 &=& 0.5 \\ z_2 &=& 4.5 \end{array}$ & $0.091$ & $0.532$ & $0.244$ \\
\hline
11 & $\frac{1+a_2}{1+a_2\exp(-a_1 z)}$ & $\begin{array} {lcl} a_1 &=& 3.4 \\ a_2 &=& 22.0 \end{array}$ & $-0.059$ & $0.372$ & $0.300$ \\
\hline
12$^b$ & $\left[ (1+z)^{a_1 a_4} + \left( \frac{1+z}{B} \right)^{a_2 a_4} + \left( \frac{1+z}{C} \right)^{a_3 a_4} \right]^{1/a_4}$ & $\begin{array} {lcl} a_1 &=& 1.6 \\ a_2 &=& -1.2 \\ a_3 &=& -5.7 \\ a_4 &=& -1.62 \\ z_1 &=& 1.7 \\ z_2 &=& 5.0 \end{array}$ & $0.058$ & $0.211$ & $0.010$ \\
\hline
13 & $ \begin{cases} (1+z)^\alpha & z \leqslant z_\star \\ (1+z_\star)^{\alpha-\beta}(1+z)^\beta & z > z_\star \end{cases} $ & $\begin{array} {lcl} \alpha &=& 2.1 \\ \beta &=& -0.7 \\ z_\star &=& 3.6 \end{array}$ & $-0.448$ & $-0.051$ & $-0.185$ \\
\hline
14$^c$ & $\frac{a_1(1+z)^{a_2}}{1+[(1+z)/a_4]^{a_3}}$ & $\begin{array} {lcl} a_2 &=& 2.7  \\ a_3 &=& 5.6  \\ a_4 &=& 2.9  \end{array}$ & $0.255$ & $0.223$ & $0.180$ \\
\hline
15$^c$ & $\frac{a_1(1+z)^{a_2}}{1+[(1+z)/a_4]^{a_3}}$ & $\begin{array} {lcl} a_2 &=& 2.14 \\ a_3 &=& 3.41 \\ a_4 &=& 3.86 \end{array}$ & $-0.198$ & $0.090$ & $0.027$
\enddata
\tablecomments{SFR1--SFR10 come from (\citealt{le17}, see references therein). SFR11 is taken from \citet{virgili09}, SFR12 from \citet{kobayashi13}, SFR13 from \citet{howell14}, SFR14 from \citet{madau14}, and SFR15 from \citet{alavi16}.\\$^a$With $a_0 = 1$, $b_0 = \frac{a_0(1+z_1)^\alpha}{(1+z_1)^\beta}$, $c_0 = \frac{(1+z_2)^\beta}{(1+z_2)^\gamma}$, where the typos in \citet{le17} were corrected.\\$^b$Where $B = \left( 1+z_1 \right)^{1-a_1/a_2}$, $C = \left( 1+z_2 \right)^{(a_2-a_2)/a_3} \left( 1+z_2 \right)^{1-a_2/a_2}$.\\$^c$Where $a_1 = 1+1/a_4^{a_3}$. }
\end{deluxetable*}

\section{Statistical Model}
\label{model}

The core idea exploited herein is that since the observed $T_{90}^{\rm obs}$ and intrinsic $T_{90}^{\rm int}$ durations are related via $T_{90}^{\rm obs} = (1+z) T_{90}^{\rm int}$, so that
\begin{equation}
\log T_{90}^{\rm obs} = \log(1+z) +\log T_{90}^{\rm int},
\label{eq7}
\end{equation}
therefore one can derive the distribution of $T_{90}^{\rm obs}$ by assuming the distributions of $z$ and $T_{90}^{\rm int}$.

As in the rest frame the durations are sufficiently well described by a mixture of two Gaussian components \citep{tarnopolski16b,tarnopolski16c,zitouni18}, they are hereinafter modeled as such. The redshift distribution is then crucial. It is modeled as
\begin{equation}
P_0(z) = N_{P_0} \frac{e(z)F(z)}{1+z},
\label{eq8}
\end{equation}
where $F(z)=\frac{\d V(z)}{\d z} = \frac{4\pi c d_p^2(z)}{H_0 h(z)}$ is the comoving volume element, with $h(z) = \sqrt{\Omega(1+z)^3+1-\Omega}$, and $d_p(z) = \frac{c}{H_0}\int_0^z\frac{\d x}{h(x)}$ is the proper distance. The latest cosmological parameters within a flat $\Lambda$CDM model \citep{Planck2018} are employed: $H_0=67.4\,{\rm km}\,{\rm s}^{-1}\,{\rm Mpc}^{-1}$ and $\Omega=0.315$. The dimensionless SFR, $e(z)$, is scaled so that $e(0)=1$. The GRB rate is expected to trace the SFR, at least up to moderate redshifts \citep{jakobsson05,schulze15}; therefore one is proportional to the other. The normalizing factor $N_{P_0}$, ensuring that $\int_0^{z_{\rm max}} P_0(z)\d z = 1$, includes this proportionality coefficient. 

Taken into account is also the Schechter luminosity function\footnote{The broken power-law model, $$\phi_z(L) \propto \begin{cases}\begin{split}\left( \frac{L}{L_b} \right)^{-\nu_1}, & \quad L\leqslant L_b \\ \left( \frac{L}{L_b} \right)^{-\nu_2}, & \quad L>L_b\end{split}\end{cases},$$ where $L_b=L_{b,0}(1+z)^\delta$, and with parameters from \citet{paul18}, was also examined, but gave very similar results, which are therefore not reported herein.} \citep{paul18}
\begin{equation}
\phi(L) = \frac{1}{L_b\Gamma(1-\nu)}\left( \frac{L}{L_b} \right)^{-\nu} \exp\left( -\frac{L}{L_b} \right),
\label{eq9}
\end{equation}
where $\nu=0.6$, $L_b = 5.4\cdot 10^{52}\,{\rm erg}\,{\rm s}^{-1}$, $\Gamma(\boldsymbol{\cdot})$ is the Euler gamma function, and $\int_0^\infty \phi(L)\d L = 1$. Considering a fiducial flux limit of the detector ($S_{\rm lim} = 3.2\cdot 10^{-8}\,{\rm erg}\,{\rm cm}^{-2}\,{\rm s}^{-1}$ for Swift, and $S_{\rm lim} = 8\cdot 10^{-8}\,{\rm erg}\,{\rm cm}^{-2}\,{\rm s}^{-1}$ for Fermi; \citealt{paul18}) leads to a redshift-dependent minimum luminosity that can be detected by a given instrument, $L_{\rm lim}(z) = 4\pi d_L^2(z) S_{\rm lim}k(z)$, where $d_L(z) = (1+z)d_p(z)$ is the luminosity distance, and the $k$-correction reads
\begin{equation}
k(z) = \frac{\int\limits_{E_1}^{E_2} ES(E)\d E}{\int\limits_{E_1(1+z)}^{E_2(1+z)} ES(E)\d E},
\label{eq10}
\end{equation}
where $S(E)$ is the \citet{band93} energy distribution, with $\alpha=-1,\,\beta=-2.3,\,E_0=511\,{\rm keV}$. For Swift/BAT, $[E_1,E_2]=[15,150]\,{\rm keV}$, and for Fermi/GBM, $[E_1,E_2]=[10,1000]\,{\rm keV}$. Therefore, a term
\begin{equation}
\int\limits_{L_{\rm lim}(z)}^\infty \phi(L)\d L = \frac{\Gamma\left( 1-\nu , \frac{L_{\rm lim}(z)}{L_b} \right)}{\Gamma\left( 1-\nu \right)}
\label{eq11}
\end{equation}
is introduced that multiplies $P_0(z)$. Here, $\Gamma(\boldsymbol{\cdot},\boldsymbol{\cdot})$ is the incomplete gamma function. 

Finally, the efficiency of Swift/BAT is\footnote{The value for $z\geqslant 5.96$ was adjusted to ensure continuity at $z=5.96$.} \citep{howell14}
\begin{equation}
\eta(z) =
\begin{cases}
\begin{split}
& -0.01+1.02\exp\left( -z/1.68 \right), & z<5.96 \\
& -0.01+1.02\exp\left( -5.96/1.68 \right), & z\geqslant 5.96
\end{split}.
\end{cases}
\label{eq12}
\end{equation}
In case of Fermi/GBM, no such straightforward expression is available; hence, the efficiency is not considered. Eventually, taking into account the detector properties, the observed redshift distribution is modeled as
\begin{equation}
P(z) = N_P \frac{e(z)F(z)}{1+z} \frac{\Gamma\left( 1-\nu , \frac{L_{\rm lim}(z)}{L_b} \right)}{\Gamma\left( 1-\nu \right)} \eta(z),
\label{eq13}
\end{equation}
where $N_P$ ensures normalization, $\int_0^{z_{\rm max}} P(z)\d z = 1$. Effectively, for Fermi simply $\eta(z)=1$ is utilized.

$P_0(z)$ can be thought of as a {\it true} redshift distribution, i.e. if not being limited by the capabilities of the detector (considering only the cosmology and SFR), which are taken into account in $P(z)$. In other words, $P_0(z)$ simulates an ideal telescope that can detect any GRB, regardless of their luminosity or distance.

\section{Results}
\label{results}

\begin{figure}
\includegraphics[width=\columnwidth]{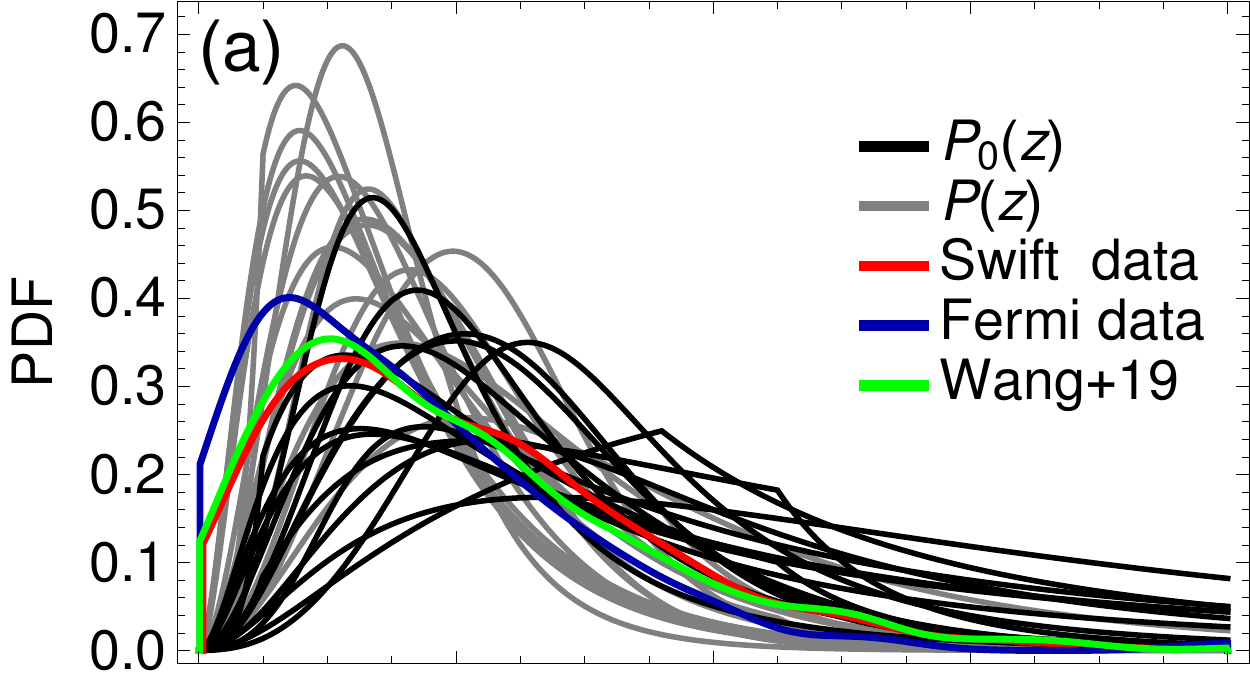}
\includegraphics[width=\columnwidth]{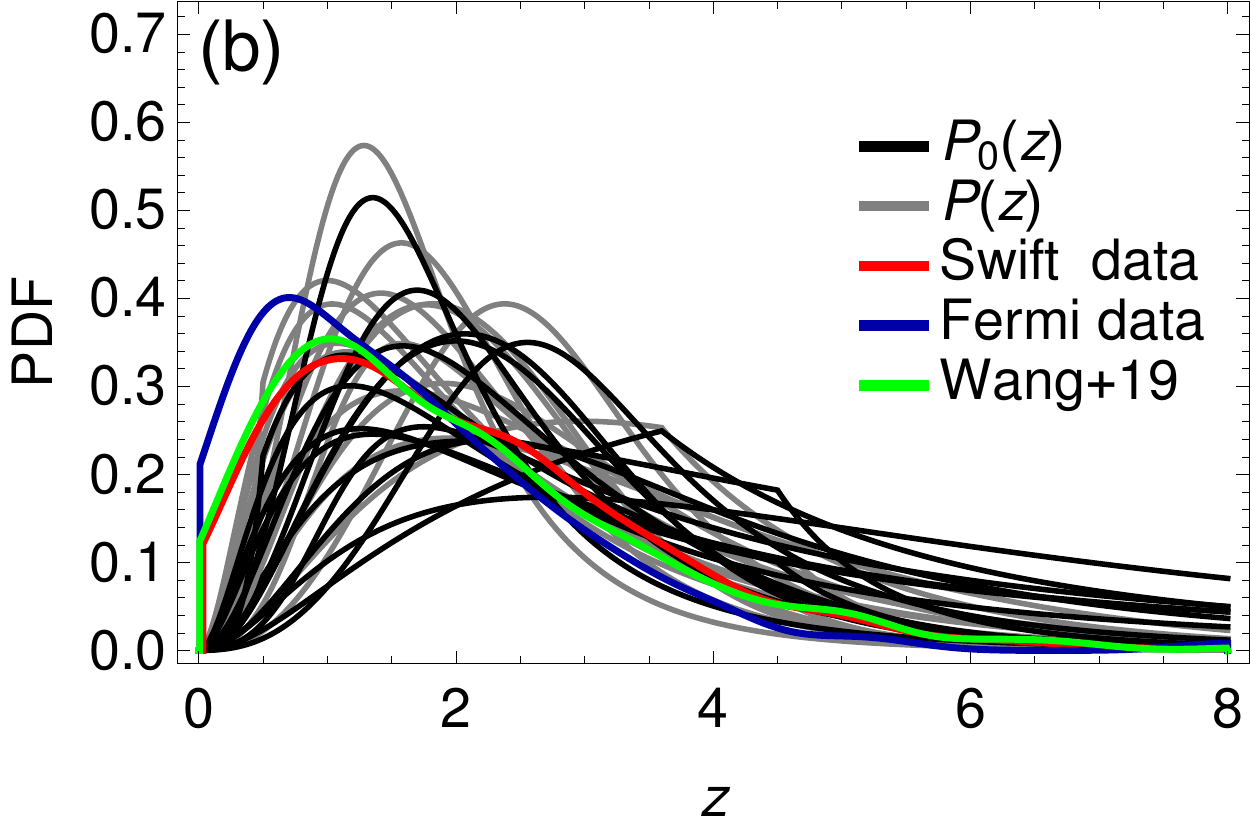}
\caption{The distributions of redshifts, modeled with various SFRs for (a) Swift and (b) Fermi detector responses (see the text for details). Also shown are the observed distributions of Swift (red line) and Fermi (blue line) samples. The green line is the sample compiled by \citet{wang20}. }
\label{fig1}
\end{figure}

The SFRs from Sect.~\ref{sfrs} were used to calculate the redshift distributions $P_0(z)$ and $P(z)$ from Equations~(\ref{eq8}) and (\ref{eq13}), respectively. They were next transformed\footnote{A standard transformation of the random variable.} to obtain the distributions of $\log(1+z)$, whose skewnesses were calculated as $\gamma=\mu_3/\sigma^3$, where $\mu_3 = E\left[ \left(X-E[X]\right)^3 \right]$ is the third central moment of a random variable $X$, and $\sigma$ is its standard deviation. The results are gathered in Table~\ref{table::sfrs}, and the resulting distributions are displayed in Fig.~\ref{fig1}, together with the actual redshift samples from Sect.~\ref{samples}. In the case of $P_0(z)$, negative skewness was attained for 10 SFRs, and SFR8 led to the most negatively skewed redshift distribution. The skewness of Swift $P(z)$ was negative only for SFR4 and SFR13. For Fermi $P(z)$, negative skewness was returned for 5 SFRs, and SFR13 turned out to lead to the most negatively skewed distribution.

Let $T,Z$ be independent random variables, corresponding to the distributions of $\log T_{90}^{\rm int}$ and $\log(1+z)$, respectively. Then the skewness of $T+Z$ (i.e., the skewness of $\log T_{90}^{\rm obs}$, according to Equation~(\ref{eq7}); see Appendix \ref{appB}) is
\begin{equation}
\gamma(T+Z) = \frac{\gamma(T)\sigma^3(T) + \gamma(Z)\sigma^3(Z)}{\left[ \sigma^2(T) + \sigma^2(Z) \right]^{3/2}}.
\label{eq14}
\end{equation}
Using Equation~(\ref{eq14}) and Swift $P(z)$, the skewness of the redshift distribution contributes nearly nothing ($0.3\%$) to the observed skewness. In the case of $P_0(z)$, i.e. in the case of an ideal detector with no flux limit and ultimate efficiency, 4\% of the observed skewness can be attributed to the cosmological dilation. In the case of Fermi, these proportions are 3\% and 12\%, respectively. Note that the former should be diminished even further after including the (unknown, but monotonically decreasing) redshift efficiency of the detector.

\section{Discussion}
\label{discussion}

\subsection{Stochastic representation}
\label{disc::5.1}

If there was a simple stochastic representation of the observed durations, as per Eq.~(\ref{eq7}), one could fit the data and infer the statistical properties of the intrinsic durations as well as of the redshifts. Assuming Gaussian distributions for $\log T_{90}^{\rm int}$ and $\log(1+z)$ would lead to a Gaussian one for $\log T_{90}^{\rm obs}$. This scenario is easily rejected, as both the observed durations and the redshifts are clearly not normal: the latter by definition, as $z>0$ demands a truncation. Likewise, one cannot consider a convolution of normal and $\mathcal{SN}$ random variables, which while also an $\mathcal{SN}$ distribution,\footnote{It is easy to prove with moment-generating functions that $\mathcal{N}(\mu_1,\sigma_1^2)+\mathcal{SN}(\mu_2,\sigma_2^2,\lambda_2)=\mathcal{SN}(\mu,\sigma^2,\lambda')$, where $\mu=\mu_1+\mu_2$, $\sigma^2=\sigma_1^2+\sigma_2^2$, and $\lambda'=\lambda_2\sigma_2/\sqrt{\sigma_1^2\left( 1+\lambda_2^2 \right)+\sigma_2^2}$. For completeness, the distribution of a sum of two $\mathcal{SN}$ random variables is expressed via Kamp\'{e} de F\'{e}riet functions \citep{nadarajah17}. }, would require the redshifts to be normal or $\mathcal{SN}$.

A promising working model might have been the sum of a normal and truncated normal random variables, as per Sect.~\ref{distributions}. The redshift distribution is in fact sufficiently well modeled with a $\mathcal{TN}$ distribution. Unfortunately, this model cannot explain the distribution of $\log T_{90}^{\rm obs}$. Note that $b>0$ and $d>0$, since they are related to standard deviations, and also $c>0$ (due to $z>0$). This means that $\lambda>0$ (see Appendix \ref{appA}), which limits the $\mathcal{SNE}$ distribution in this setting to be positively skewed, while previous works \citep{tarnopolski16a,tarnopolski16c,kwong} and the $\mathcal{SN}$ fitting results from Sect.~\ref{fitting} imply that the duration distribution is negatively skewed.

In fact, assuming a normal distribution for $\log T_{90}^{\rm int}$, as per Eq.~(\ref{eq14}) the skewness of the redshift distribution needs to be negative to produce negative skewness in the observed durations, while overall it is clearly positive (although some SFR models are able to produce a small to moderate negative skewness). Therefore, no such simple statistical representation can model both duration distributions (observed and intrinsic) and the redshift distribution. Note that the intrinsic durations examined herein are actually well described by an $\mathcal{SN}$ distribution as well. The intrinsic skewness might then lead to the observed skewness. A bigger homogenous sample, however, is needed to robustly assess this possibility.

Finally, an effect contrary to the one presumed (and rejected) herein was observed for single-pulse GRBs: it turns out that only the brightest portions of such light curves of distant GRBs are observable on Earth \citep{kocevski13}. This is due to an increasing with redshift signal-to-noise ratio, which leads to obscuring larger portions of the light curve for greater redshifts. Therefore, the sensitivity of the detector leads to a decrease of duration of distant GRBs. Such an effect can overwhelm the influence of the cosmological dilation. Additionally, the $k$-correction also affects the observed duration, since the low-energy region of the GRB spectrum becomes invisible to the detector when the redshift increases. This means that the same burst observed at different redshifts would have both a different duration and spectrum. However, for multi-pulse GRBs this might be different, since the quiescent phase between the pulses would be subject to the cosmological dilation, leading to longer durations of distant GRBs. Weak evidence for such phenomena have been reported \citep{zhang13,littlejohns14,turpin16}. Overall, it appears that neither the cosmological redshift nor the instrument properties can explain the high skewness of the observed durations.

\subsection{Physical Models}
\label{disc::5.2}

A straightforward explanation for the observed skewness was put forward by \citet{zitouni15}, who suggested that the distribution of the envelope masses of the progenitors of long GRBs might be inherently skewed. The simple reasoning is that lighter progenitor stars will eject lighter envelopes, which are the {\it fuel} for the GRBs, and that their freefall times will allow the accretion disks to be powered for a limited period of time. As there are more relatively light stars than massive ones, this seems a plausible possibility. Its testing might be done with a stellar population synthesis in order to gather an ensemble of progenitor stars, evolving them with a stellar evolution code \citep{paxton18}, and retrieving the envelope masses at the final stages of the stars' lives.

A next step might be modeling the emerging relativistic jet that powers the GRB. With such an approach it was already shown that blue supergiants can be the source of ultra-long GRBs (ULGRBs), i.e. those with $10^3\,{\rm s}\lesssim T_{90}\lesssim 10^4\,{\rm s}$ \citep{perna18}, and that Wolf--Rayet stars lead to time scales reflecting the durations of regular long GRBs. The precise duration of ULGRBs depends on the viewing angle, $\theta$, which suggests that for regular long GRBs the distribution of $\theta$ also plays a role in shaping the distribution of $T_{90}$.

\section{Conclusions}
\label{summary}

A null hypothesis that the observed skewness of the long GRBs' duration distribution is predominantly caused by the cosmological dilation was tested. Several forms for the SFR (proportional to the GRB rate) were considered. It was found that the convolution with the redshift distribution can explain only up to 12\% of the observed negative skewness, while in several cases the resulting skewness was positive, inconsistent with observational data. When detector properties (flux limit, $k$-correction, and, in the case of Swift, the redshift efficiency) were taken into account, the skewness diminished even further. Therefore, neither the cosmological dilation nor the detector responses are able to explain the skewness of the duration distribution. Its origin needs to be searched for in the physical processes governing the progenitors of GRBs.

\acknowledgments
The author acknowledges support by the Polish National Science Center through the OPUS grant No. 2017/25/B/ST9/01208.

\newpage

\appendix
\section{PDF of the $\mathcal{SNE}$ distribution}
\label{appA}

Let $U,V$ be independent random variables such that $U\sim \mathcal{TN}_{[0,+\infty)}(c,d)$, $V\sim \mathcal{N}(a,b)$. Their PDFs are
\begin{equation}
f^{(\mathcal{TN})}(u)=\frac{1}{\sqrt{2\pi}d\Phi\left(\frac{c}{d}\right)}\exp\left[-\frac{(u-c)^2}{2d^2}\right],\quad u>0
\label{eqA1}
\end{equation}
\begin{equation}
f^{(\mathcal{N})}(v)=\frac{1}{\sqrt{2\pi}b}\exp\left[-\frac{(v-a)^2}{2b^2}\right],\quad -\infty<v<+\infty
\label{eqA2}
\end{equation}
Define $U_1,U_2$ such that $U=U_1-U_2$, $V=U_2$; thence the Jacobian $|J|=1$, and the joint PDF is $f_{U_1,U_2}(u_1,u_2) = f^{(\mathcal{TN})}\left[ u(u_1,u_2) \right] f^{(\mathcal{N})}\left[ v(u_1,u_2) \right] |J|$. The marginal distribution of $U_1$:
\begin{equation}
\begin{split}
f_{U_1=V+U}(u_1) &= \int\limits_{-\infty}^{+\infty} f^{(\mathcal{TN})}\left[ u(u_1,u_2) \right] f^{(\mathcal{N})}\left[ v(u_1,u_2) \right] |J| \d u_2 \\
 &= \frac{1}{\sqrt{2\pi}b}\frac{1}{\sqrt{2\pi}d}\frac{1}{\Phi\left(\frac{c}{d}\right)} \int\limits_{-\infty}^{+\infty} \exp\left[ -\frac{(u_2-a)^2}{2b^2} \right] \exp\left[ -\frac{(u_1-u_2-c)^2}{2d^2} \right] \mathbbm{1}_{u_1>u_2} \d u_2 \\
 &= \frac{\varphi\left( \frac{u_1-\mu}{\sigma} \right) \Phi\left( \lambda\frac{u_1-\mu}{\sigma}+\xi \right)}{\sigma\Phi\left( \frac{\xi}{\sqrt{1+\lambda^2}} \right)} = f^{(\mathcal{SNE})}(u_1),
\end{split}
\label{eqA3}
\end{equation}
where the indicator function $\mathbbm{1}_{u_1>u_2}$ changes the limits of integration from $(-\infty,+\infty)$ to $(-\infty,u_1)$, and $(\mu,\sigma,\lambda,\xi)$ are given by
\begin{equation}
\begin{cases} 
      \mu = a+c \\
      \sigma = \sqrt{b^2+d^2} \\
      \lambda = \frac{d}{b} \\
      \xi = \frac{c}{bd}\sqrt{b^2+d^2}
\end{cases}
,
\label{eq6}
\end{equation}
hence meaning that $U_1\sim \mathcal{SNE}(\mu,\sigma,\lambda,\xi)$.

\section{Skewness of a sum of random variables}
\label{appB}

Consider
\begin{equation}
\gamma(X+Y) = \frac{\mu_3(X+Y)}{\left({\rm var}(X+Y)\right)^{3/2}},
\label{eqB1}
\end{equation}
where, after some standard manipulations,
\begin{equation}
\begin{split}
\mu_3(X+Y) &= E\left[\left( X+Y-E(X+Y) \right)^3\right] \\
 &= \mu_3(X)+\mu_3(Y)+3{\rm cov}(X^2,Y)+3{\rm cov}(X,Y^2)-6\left( E[X]+E[Y] \right){\rm cov}(X,Y),
\end{split}
\label{eqB2}
\end{equation}
and since ${\rm var}(X+Y) = {\rm var}(X)+{\rm var}(Y)+2{\rm cov}(X,Y)$, one obtains the skewness after substituting into Eq.~(\ref{eqB1}). Setting the covariances to zero yields
\begin{equation}
\gamma(X+Y) = \frac{\mu_3(X)+\mu_3(Y)}{\left( {\rm var}(X)+{\rm var}(Y) \right)^{3/2}},
\label{eqB3}
\end{equation}
which upon inserting $\mu_3(X) = \gamma(X) \left( {\rm var}(X) \right)^{3/2} \equiv \gamma(X) \sigma^3(X)$, and likewise for $Y$, results in Eq.~(\ref{eq14}).

\bibliography{bibliography}{}
\bibliographystyle{aasjournal}

\end{document}